\documentclass[%
 reprint,
 amsmath,amssymb,
 aps,
]{revtex4-1}

\usepackage{graphicx}% Include figure files
\usepackage{dcolumn}% Align table columns on decimal point
\usepackage{bm}% bold math
\newcommand*{\affaddr}[1]{#1} % No op here. Customize it for different styles.
\newcommand*{\affmark}[1][*]{\textsuperscript{#1}}

\begin{document}

\preprint{APS/123-QED}

\title{Probing the Stau-Neutralino Coannihilation Region at the LHC with a soft tau lepton and an ISR jet} % Force line breaks with \\

%\author[*]{Andr\'es Fl\'orez}
%\author[**]{Alfredo Gurrola}
% \author[*]{Carlos \'Avila}

%\affiliation[*]{Universidad de los Andes}
%
%\affiliation[**]{Vanderbilt University}

\author{
Andr\'es Fl\'orez\affmark[1], Luis Bravo\affmark[1], Alfredo Gurrola\affmark[2], Carlos \'Avila\affmark[1], Manuel Segura\affmark[1], Paul Sheldon\affmark[2] and Will Johns\affmark[2]\\
\affaddr{\affmark[1] Physics Department, Universidad de los Andes, Bogot\'a, Colombia}\\
\affaddr{\affmark[2] Department of Physics and Astronomy, Vanderbilt University, Nashville, TN, 37235, USA}\\
%\email{\{A,B,C,D,E\}@university.edu}\\
%\affaddr{\LaTeX\ University}
}

%\author{Carlos \'Avila}
%\affiliation {Universidad de los Andes}
%\author{Luis Bravo}
%\affiliation {Universidad de los Andes}
%\author{Manuel Segura}
%\affiliation {Universidad de los Andes}

%Andr\'es Flo\'orez$^{1}$, Alfredo Gurrola$^{2}$
%
%Universidad de los Andes$^{1}$, Vanderbilt University$^{2}$


%\affiliation{%

%\affiliation{
%\affiliation{
%\affiliation{%

\date{\today}% It is always \today, today,

\begin{abstract}

We present a feasibility study, to search for dark matter at the LHC, in events with one soft hadronically decaying tau lepton and missing transverse energy recoiling against a hard $p_{T}$ jet from initial state radiation. This methodology allows the search for Supersymmetry   in compressed mass spectra regions, where the mass difference between the lightest neutralino, $\tilde\chi_1^0$, and the stau (the tau superpartner), $\tilde{\tau}$, is small. Several theoretical models predict a direct connection between thermal Bino dark matter and staus within this scenario.  We show that compressed regions, not excluded by ATLAS nor CMS experiments, are opened up with the increase in experimental sensitivity reached with the proposed methodology.  
The requirement of a hard jet from initial state radiation combined with a soft tau lepton is effective in reducing Standard Model backgrounds, providing expected significances greater than 3$\sigma$ for $\tilde{\chi}_{1}^{\pm}$ masses up to 300 GeV and $\tilde{\tau}$-$\tilde{\chi}_{1}^{0}$ mass gaps below 25 GeV with only 30 fb$^{-1}$ of 13 TeV data from the LHC.

%and that the mass difference between the $\tilde{\tau}$ and the $\tilde{\chi}^{0}_{1}$ is small.

%\begin{description}
%\end{description}
\end{abstract}

\pacs{Valid PACS appear here}% PACS, the Physics and Astronomy
\maketitle

%\tableofcontents

\section{\label{sec:level1}Introduction}

The identity of Dark Matter (DM) is one of the most interesting and relevant topics in particle physics today. Currently, there are several direct and indirect searches for DM performed by different experiments, 
such as superCDMS \cite{Agnese:2015nto}, LZ \cite{Akerib:2015cja},  AMS2 \cite{Accardo:2014lma}, ATLAS \cite{Aad:2008zzm} and CMS \cite{Chatrchyan:2008aa}, among others. These experiments are  trying to find evidence of the existence of DM particles motivated by hypothetical models, in some cases, or by indirect cosmological observations. Nevertheless, there is no conclusive evidence thus far that could 
shed some light on the particle nature of DM.

At the CERN LHC accelerator, the ATLAS and CMS experiments have an extensive physics program to search for DM, especially in new physics models of   
Supersymmetry (SUSY)\cite{SUSY1, SUSY2, SUSY3, SUSY4, SUSY5}, which resolves many problems inherent in the Standard Model (SM) and naturally provides a DM candidate in the form of the lightest neutralino ($\tilde{\chi}_{1}^{0}$). A broad set of final states have been used to probe the $\tilde{\chi}_{1}^{0}$ using cascade decays of heavier colored and electroweak SUSY particles  
\cite{ATLAS1, ATLAS2, ATLAS3, CMS1, CMS2}. 
The production of these DM candidates has been excluded, by both experiments, for $\tilde{\chi}_{1}^{0}$ masses that range from 100 GeV to roughly 800 GeV, depending on the  final state studied and on the physics model used to interpret the data. Nevertheless, compressed mass spectra regions, where the mass difference $\Delta m$ between the heavier SUSY particles and the $\tilde{\chi}_{1}^{0}$ is small, are very difficult to probe at the 
LHC, due to constrains driven by the ability to trigger, with low enough rate, on events containing low $p_{T}$ objects in addition to experimental difficulties involved with identifying them with high enough efficiency amongst the large hadronic activity associated with a proton-proton collider. 
For example, searches for chargino ($\tilde{\chi}_{i}^{\pm}$) and neutralino ($\tilde{\chi}_{j}^{0}$) production in final states with one or more leptons and missing transverse momentum exhibit limited sensitivity to models with SUSY particles that decay predominantly to $\tau$ leptons, with exclusion limits of $\approx 100$ GeV for $\Delta m < 50$ GeV, due to the larger backgrounds associated with $\tau$ lepton reconstruction.

The main focus of this letter is to propose a new search at the LHC to target compressed mass spectra regions in the electroweak sector, in models which predominantly produce $\tau$ leptons, where the current experimental sensitivity is limited. The study of compressed $\tilde{\tau}$'s is of 
special interest in thermal Bino DM cosmology models considering $\tilde{\tau} - \tilde{\chi}^{0}_{1}$  
co-annihilation, as it is proposed in several papers ~\cite{DMmodels1,  DMmodels3}, in order to obtain the correct relic DM density observed today. 

The use of Vector Boson Fusion (VBF) topologies to target difficult compressed mass spectra scenarios for the production of SUSY with $\tilde{\tau}$'s, has been proposed as a new experimental handle at the LHC \cite{VBF1}. 
This search has been recently published by CMS \cite{VBF2}, showing better sensitivity in very compressed regions with respect to previous searches by ATLAS and CMS \cite{Mono1, Mono2}. 
Although VBF is a good tool to probe compressed spectra and DM \cite{DMmodels2}, with better signal-to-background ratios due to its rejection power for QCD processes, the small VBF signal cross-sections motivate us to find a complementary method with higher production rate, which translates in less luminosity needed for a potential discovery in the short term. 
We propose a complementary handle to target compressed staus, searching for the production of one hadronic $\tau$ lepton ($\tau_{h}$ ) and at least one high $p_{T}$ jet 
from initial state radiation (ISR).  

The SUSY $\tilde{\tau}$'s can be produced directly in pairs or through cascade decays of the lightest chargino, $\tilde{\chi}^{\pm}_{1}$, and  the next-to-lightest neutralino, 
$\tilde{\chi}^{0}_{2}$, in processes such as $\tilde{\chi}^{\pm}_{1} \tilde{\chi}^{\mp}_{1} \to \tilde{\tau} \tilde{\tau} \nu_{\tau} \nu_{\tau}$, $\tilde{\chi}^{0}_{2}\tilde{\chi}^{0}_{2} \to \tilde{\tau}\tilde{\tau} \tau \tau$, $\tilde{\chi}^{\pm}_{1}\tilde{\chi}^{0}_{2} \to  \tilde{\tau} \nu_{\tau} \tilde{\tau} \tau $ and $\tilde{\chi}^{\pm}_{1}\tilde{\chi}^{0}_{1} \to \tilde{\tau} \nu_{\tau} \tilde{\chi}^{0}_{1}$. 
Hadronic decays of $\tau$ leptons have the largest branching fraction and thus final states with a $\tau_{h}$ provide the best experimental sensitivity.

While the above processes result in final states with multiple $\tau$ leptons, the compressed mass spectra scenario of interest in this paper results in low $p_{T}$ visible decay products,  making it difficult to reconstruct and identify multiple $\tau$ leptons.
Furthermore, semi-leptonic decays of $\tau$ leptons result in lower average $p_{T}$ than hadronic decays, while also being largely indistinguishable from prompt production of electrons and muons. 

Therefore, the above characteristics motivate us to focus on events with one $\tau_{h}$ candidate. 
Similar to the monojet searches, the use of a high $p_{T}$ ISR jet in the event topology
is expected to create a recoil effect that facilitates both, the detection of missing 
transverse momentum in the event ($p^{miss}_{T}$), and the identification of the soft $\tau_{h}$ due to the natural kinematic boost. Additionally, the inclusion of a high $p_{T}$ jet in the event topology provides an experimental handle to trigger on these type of events with soft $\tau_{h}$ candidates.

\section{Samples and simulation}

Signal and background samples were simulated using an interface between MadGraph (v2.2.3) \cite{MADGRAPH} for the events generation, PYTHIA (v6.416) \cite{Sjostrand:2006za} for the hadronization process and Delphes (v3.3.2) \cite{deFavereau:2013fsa} to include detector effects. The main background sources come from the production of Z and W vector bosons with associated jets, referred to as Z+jets and W+jets.
Background events with up to two associate jets were generated. Jet merging and matching was performed based on the MLM algorithm \cite{MLM}. This algorithm requires the optimization of two variables related to the jet definition, qcut and xqcut. The xqcut is defined as the minimal distance required among partons at MadGraph level. The qcut is a measure of the minimum energy spread for a clustered jet in PYTHIA.  
The optimization is performed by studying the differential jet rate distribution until obtaining a smooth transition between 
events with zero and one jets, and between events with one and two jets. The optimal values found for our simulations were a xqcut of 15 for both  backgrounds, and a 
qcut of 35 GeV for  Z+jets and 30 GeV for W+jets.
At generation level, leptons were required to have a $p_{T} (\ell) > 10$ GeV and $|\eta (\ell)| < 2.5$, while jets are required to have a minimum $p_{T}$ threshold of 20 GeV and $|\eta| < 5.0$. 
For Z+jets events, an additional constrain on the reconstructed mass of the two leptons was applied, in order to suppress events with masses below 50 GeV. 

The signal samples were produced considering two cases in the context of the R-parity conserving Minimal Supersymmetric Standard Model (MSSM).
The first case considered direct production of $\tilde{\tau}$ pairs and an ISR jet and the second case included additional production of 
$\tilde{\tau}$ events through cascade decays of $\tilde{\chi}^{\pm}_{1}$ or $\tilde{\chi}^{0}_{2}$. The benchmark signal samples were produced under the three following assumptions. First, the $\tilde{\chi}^{\pm}_{1}$ and the  $\tilde{\chi}^{0}_{2}$ are wino-like and mass degenerate, while the $\tilde{\chi}_{1}^{0}$ is mostly Bino. Second, we considered only scenarios  where  the mass difference between the $\tilde{\tau}$ and the $\tilde{\chi}^{0}_{1}$ is always less or equal to 25 GeV, aimed at the $\tilde{\tau}$-$\tilde{\chi}_{1}^{0}$ co-annihilation region, and with mass equal to $m(\tilde{\tau})=0.5m(\tilde{\chi}_{1}^{\pm})+0.5m(\tilde{\chi}_{1}^{0})$. Finally, we studied regions where the mass difference between the $\tilde{\chi}^{0}_{1}$ and the $\tilde{\chi}_{1}^{\pm}$ is below 50 GeV, in order to study areas  of the SUSY phase space where the ATLAS and CMS searches have limited experimental sensitivity. We scaned the regions of interest using $\tilde{\chi}^{\pm}_{1}$ masses ranging from 100 GeV to 400 GeV, in steps of 100 GeV, and $\Delta m(\tilde{\tau}, \tilde{\chi}^{0}_{1})$ from 5 GeV to 25 GeV, in steps of 5 GeV.

\section{Event selection criteria}

\begin{figure}
 \begin{center} 
 \includegraphics[width=0.45\textwidth, height=0.35\textheight]{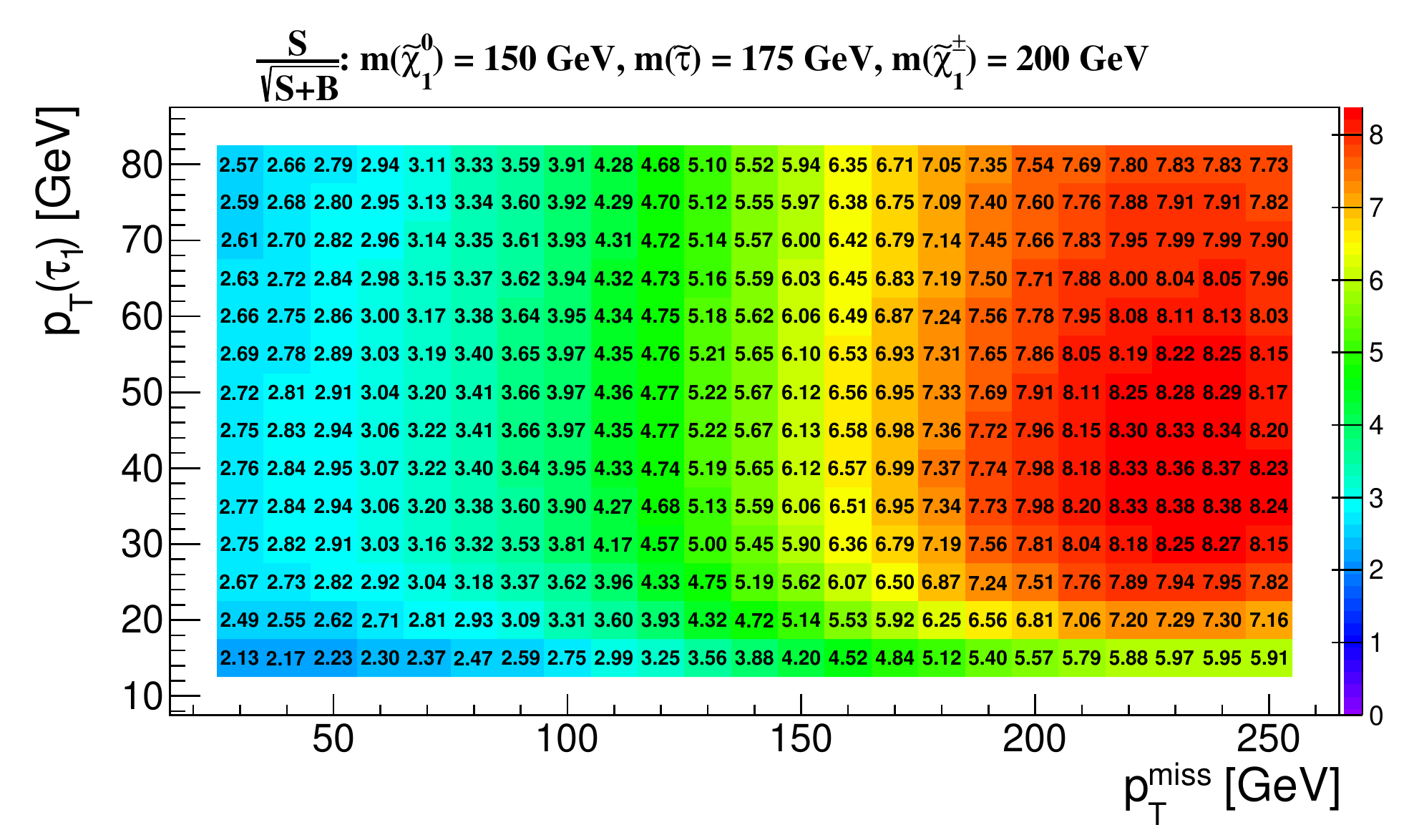}
 \end{center}
 \caption{Results of the $p_{T}^{max}(\tau_{h})$ vs. $p_{T}^{miss}$ optimization process, targeting best significance $S/\sqrt{S+B}$, using events selected
with $p_{T}^{lead}(jet) > 100$ GeV, $p_{T}(\tau_{h}) > 15$ GeV, and satisfying the extra lepton and b-jet vetoes. The benchmark signal point used was 
 $m(\tilde{\chi}^{0}_{1}) = 150$ GeV, $m(\tilde{\chi}^{\pm}_{1}) = 200$ GeV and $m(\tilde{\tau}) = 175$ GeV.}
 \label{fig:significance}
 \end{figure}

The event selection criteria used in the analysis is summarized in Table~\ref{tab:selections}. 
The $p_{T}$ threshold for the highest $p_{T}$ jet ($p^{lead}_{T}(jet)$) was defined through an optimization process, based on the $S/\sqrt{S+B}$ figure of merit, to estimate the signal significance. 
The $p^{lead}_{T} (jet)$ selection was also chosen to provide an experimental handle to trigger on these types of events at ATLAS and CMS. 
In order to focus on events where the ISR jet can naturally  boost the $p^{miss}_{T}$ in the opposite direction, jets are constrained to be within the tracker acceptance region, $|\eta_{jets}| <$2.5. 
We selected the highest $p_{T}$ jet in the event, as the ISR jet. The highest $p_{T}$ jet correctly identifies the ISR jet with greater than 95\% accuracy.
Events containing an isolated electron or muon, with $p_{T} > 20$ GeV, have been removed in order to suppress the contribution from the
W+jets, Z+jets  and $t\bar{t}$ backgrounds. The contribution from di-boson events, is heavily suppressed after vetoing events with two or more leptons. Events with top quarks become negligible after vetoing jets, tagged as bottom quarks, with $p_{T} > 20$ GeV and $|\eta| <$ 2.5.  
Events are required to have exactly one $\tau_{h}$ with $15 < p_{T}(\tau_{h}) < 35$ GeV and $|\eta(\tau_{h})| < 2.3$.
The selection criteria on the pseudo-rapidity of $\tau_{h}$, $|\eta(\tau_{h})| <$ 2.3, is also motivated by the geometric acceptance of the tracker sub-detectors in both experiments 
and the isolation cones placed around the $\tau_{h}$ candidates, commonly used to 
reject jets from QCD processes that can mimic the signature of a $\tau_{h}$. Jets and $\tau_{h}$ candidates passing 
the outlined selection criteria are required to be well separated in $\eta - \phi$ space by a cut of $\Delta R (\tau_{h}, jet) = \sqrt{\Delta \phi^2 + \Delta \eta^{2}}$ greater than 0.4. 
The $p_{T}(\tau_{h})$ and  $p^{miss}_{T}$ thresholds were optimized in a two dimensional plane, after passing the selection criteria described above, 
allowing us to find the most suitable combination of the two variables. The signal benchmark sample with $m(\tilde{\chi}^{0}_{1}) = 150$ GeV, $m(\tilde{\chi}^{\pm}_{1}) = 200$ GeV and $m(\tilde{\tau}) = 175$ GeV, was 
used for the optimization. The best significance is achieved when $p_{T}(\tau_{h})$  is within the range 15 GeV $ < p_{T}(\tau_{h}) < $ 35 GeV, with a $p_{T}^{miss}$ requirement above 230 GeV. After requiring a $p_{T}^{miss}$ threshold of 230 GeV, the contribution from QCD events becomes negligible.
Figure \ref{fig:significance} shows the results of the $p_{T}^{max}(\tau_{h})$ vs. $p_{T}^{miss}$ optimization process, using events selected with $p_{T}^{lead}(jet) > 100$ GeV, $p_{T}(\tau_{h}) > 15$ GeV, and satisfying the extra lepton and b-jet vetoes. The increase in signal significance due to the requirement of a soft $\tau_{h}$, as shown in Figure \ref{fig:significance}, highlights the importance of having good $\tau_{h}$ identification at low $p_{T}$. On average, a 20\% improvement in the overall signal significance for very compressed $\tilde{\tau} -\tilde{\chi}_{1}^{0}$ scenarios, is observed by lowering the $p_{T}(\tau_{h})$ threshold from 20 GeV to 15 GeV. 
  
Other sets of topological variables were analyzed, such as different combination of variables related to the angular difference in the $\phi$ plane between the highest $p_{T}$ jet, the $\tau_{h}$ and the 
$p^{miss}_{T}$. However, no significant additional discrimination between the hypothetical signal and the background was observed. Similarly, other distributions such as the scalar 
sum of the  $p_{T}$ of the jets in the event ($H_{T}$) and the ratio of $p_{T}^{miss}$ to  $p^{lead}_{T}(jet)$, $R_{m}$ \cite{RM}, were also studied. Neither the  $H_{T}$ nor the $R_{m}$ variables 
yield additional signal to background discrimination.
   
\begin{table}
\begin{center}
\caption {Cuts used to select events with one $\tau_{h}$ and at least one high $p_{T}$ ISR jet. The highest $p_{T}$ jet is tagged as the ISR jet.}
\label{tab:selections}
\begin{tabular}{ l  c }\hline\hline

Criterion & Selection \\

 \hline
 $N(e/\mu)$ & 0 \\
 $N(\tau_{h})$ & 1 \\
 $|\eta(\tau_{h})|$ &  $< 2.3$\\
  N(b-jets)  & 0\\
 $p^{lead}_{T}(jet)$ & $100$ GeV \\
  $\Delta R(\tau_{h}, jet)$ & $> 0.4$ \\
 $|\eta (jet)|$ & $< 2.5$ \\
  $p_{T}(\tau_{h})$ & $> 15$ GeV \& $< 35$ GeV\\
  $p^{miss}_{T}$ & $> 230$ GeV \\
   \hline\hline

 \end{tabular}
\end{center}

\end{table}

The transverse mass distribution between the $\tau_{h}$ and the $p^{miss}_{T}$, defined in Equation~\ref{eq:mT}, is proposed as the main signal to background discriminant, to search for a possible broad enhancement of signal events in the tails of the distributions that would indicate the presence of new physics at the LHC. 

\begin{equation}
   m_{T}(\tau_{h}, p_{T}^{miss}) = \sqrt{2p^{miss}_{T} p_{T}(\tau_{h}) (1 - cos\Delta\phi(\tau_{h}, p^{miss}_{T}))}
  \label{eq:mT}
  \end{equation}

\begin{figure}
 \begin{center} 
 \includegraphics[width=0.5\textwidth, height=0.35\textheight]{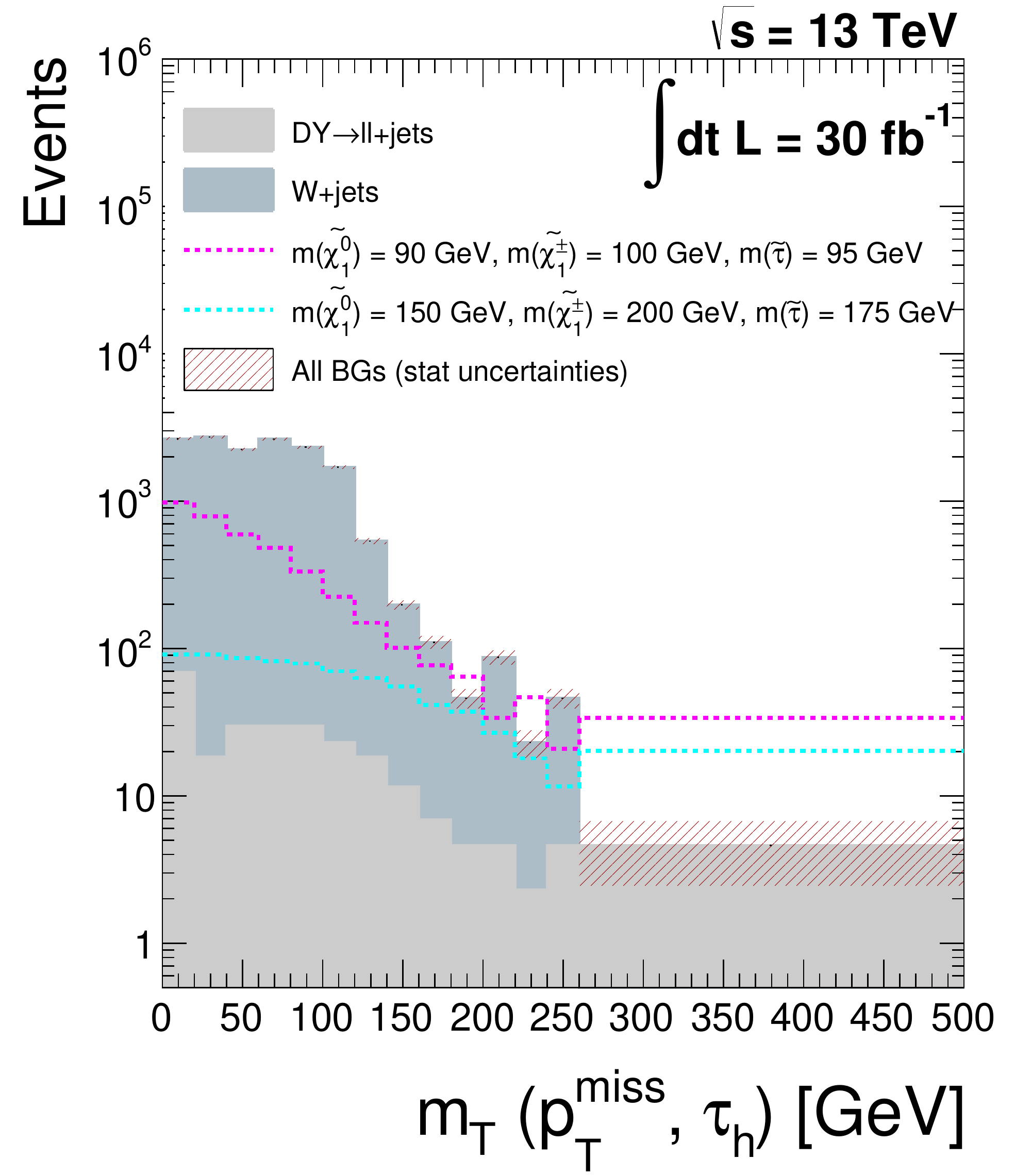}
 \end{center}
 \caption{$m_{T}(\tau_{h}, p_{T}^{miss})$ distribution for the main backgrounds and two chosen signal benchmark points, after applying the final event selection criteria.}
 \label{fig:mT}
 \end{figure}

Figure \ref{fig:mT} shows the  $m_{T}(\tau_{h}, p_{T}^{miss}) $ distribution for the main backgrounds and two different signal points, after applying all the event selection criteria outlined in Table \ref{tab:selections}. The backgrounds are stacked on top of each other while the signal is overlaid with the expected background yields. The bulk of the background distribution resides at low $m_{T}$, while the signal begins to dominate in the tails of the distribution (e.g. $m_{T} \sim 175$ GeV for the benchmark signal samples shown in Figure \ref{fig:mT}). 

Tables \ref{tab:eficiencia1} and \ref{tab:eficiencia2} present the change in the production cross section and the relative efficiencies for two signal points and the main backgrounds, after each step of 
the event selection criteria. The tables also contain information for different $m_{T}$ cuts, aimed to help the reader determine the sensitivity of the analysis.

\begin{table}

\centering
{\tiny 
\caption{{\small Change in the production cross section and relative percentage efficiencies after each step of the event selection criteria, for two signal points. The cross sections are in femtobarns and the relative efficiencies are presented in parenthesis: $\sigma (\varepsilon)$.
Signal 1 corresponds to $m(\tilde{\chi}^{0}_{1}) = 150$ GeV, $m(\tilde{\chi}^{\pm}_{1}) = 200$ GeV, $m(\tilde{\tau}) = 175$ GeV and signal 2 to $m(\tilde{\chi}^{0}_{1}) = 90$ GeV, $m(\tilde{\chi}^{\pm}_{1}) = 100$ GeV, $m(\tilde{\tau}) = 95$ GeV.}}
\label{tab:eficiencia1}
\begin{tabular}{c c c}
\hline  \hline
Criterion                                                 & Signal 1 &  Signal 2 \\\hline
No cuts                                                  & 1331.0    & 10700.0\\
$p_{T} (j_{1}) >$ 20 GeV                      &  1292.4 (97.1)    & 9854.7 (92.1)    \\ 
N(e/$\mu$) = 0                                      &  1023.6 (79.2)   &  9844.8 (99.9)     \\
N($\tau_{h}$) = 1                                  &   237.5 (23.2)    &   2480.9 (25.2)   \\
$|\eta(\tau_{1})| <$ 2.3                          &  234.9 (98.9)     &   2441.2 (98.4) \\ 
N(b-jets) = 0                                          &  215.8 (91.9)     &  2282.5 (93.5)  \\ 
$p_{T} (j_{1}) >$ 100 GeV                     &  132.1 (61.2)    &   808.0 (35.4) \\ 
$|\eta(\j_{1})| <$ 2.5                               &  128.3 (97.1)    &  776.5 (96.1) \\ 
15 GeV $< p_{T} (\tau_{h}) < $ 35 GeV &  59.9 (46.7)      &  285.0 (36.7) \\ 
$\vec{E}_{T}^{miss} >$ 230 GeV           &  25.9 (43.2)      &131.4 (46.1)\\ 
$m_{T}\,>$ 150 GeV                              &  6.1 (23.6)        & 11.0  (8.4) \\                                  
$m_{T}\,>$ 200 GeV                              &  2.6 (9.9)        &  4.5 (3.4) \\                                
$m_{T}\,>$ 250 GeV                              &  0.8 (3.1)          &1.4 (1.1) \\  \hline \hline                      
\end{tabular}
}
\end{table}

\begin{table}
{\tiny
\centering
\caption{Change in the production cross section and relative percentage efficiencies after each step of the event selection criteria, for the main backgrounds. The cross sections are in femtobarns and the relative efficiencies are presented in parenthesis: $\sigma (\varepsilon)$.}
\label{tab:eficiencia2}
\begin{tabular}{c c c}
\hline  \hline
Criterion                                                      &   DY+jets            & W+jets \\\hline
No cuts                                                   & 2240000.0             & 31800000.0 \\
$p_{T} (j_{1}) >$ 20 GeV                       &  1019200 (45.5)     &  11543400.0 (36.3)      \\
N(e/$\mu$) = 0                                      & 354681.6 (34.8)     &  7122277.8 (61.7)       \\ 
N($\tau_{h}$) = 1                                   & 148966.3 (42.0)     &  1403088.7 (19.7)        \\ 
$|\eta(\tau_{1})| <$ 2.3                           & 148519.4 (99.7)     & 1391864.0 (99.2)       \\ 
N(b-jets) = 0                                           & 143172.7 (96.4)     &  1333405.7 (95.8)   \\
$p_{T} (j_{1}) >$ 100 GeV                      & 7158.6 (5.0)          &  90671.6 (6.8)        \\ 
$|\eta(\j_{1})| <$ 2.5                                &  6764.9 (94.5)       &  84415.2 (93.1)        \\ 
15 GeV $< p_{T} (\tau_{h}) < $ 35 GeV &  1928.0 (28.5)        &  28532.3 (33.8)        \\ 
$\vec{E}_{T}^{miss} >$ 230 GeV           &  9.6 (0.5)                &  513.6 (1.8)         \\ 
$m_{T}\,>$ 150 GeV                              &  1.2 (12.5)               & 11.29 (2.2)      \\                                 
$m_{T}\,>$ 200 GeV                              &  0.6 (6.25)               & 5.29  (1.0)     \\                                 
$m_{T}\,>$ 250 GeV                              &  0.3 (3.6)                 &  0.0 (0.0)        \\    \hline  \hline                             
\end{tabular}
}
\end{table}

\section{Results}

The proposed shape based analysis of the $m_{T}$ distribution is performed using a binned likelihood following the test statistic based on the profile likelihood ratio, using the ROOTFit \cite{ROOTFit} toolkit. 
As can be seen from Figure \ref{fig:mT}, the signal sensitivity with the integrated luminosity considered is dominated by the signal and background yields in the tails of the $m_{T}$ distribution, where statistical uncertainties are expected to be more important than systematic uncertainties. However, since the proposed search strategy entails a fit of the entire $m_{T}$ distribution, it is appropriate to consider reasonable experimental systematic uncertainties to calculate projected significance as this fitting procedure can have important correlations to the background and signal uncertainties at low $m_{T}$, where statistical uncertainties are small.  The dominant sources of systematics are expected to be the uncertainty on $\tau_{h}$ identification (6\% based on \cite{CMSTauID}), $p_{T}^{miss}$ trigger efficiency (1\% based on \cite{EXO12048}), modeling of ISR (5\% based on \cite{EXO12048}), pileup effects, and the uncertainty on transfer factors used to estimate the backgrounds. While it is beyond the scope of this paper to perform studies on background estimation methods, we refer to the monojet searches with 8 TeV data \cite{EXO12048} as a reasonable choice for the uncertainty on the transfer factors used to estimate backgrounds ($\sim 5.1$\% for $p_{T}^{miss} > 250$ GeV). Therefore, a 10\% total systematic uncertainty on the signal and background yields is a reasonable choice.  In our studies, the systematic uncertainties are incorporated via nuisance parameters following the frequentist approach. A local p-value is calculated as the probability under a background only hypothesis to obtain a value of the test statistic as large as that obtained with a signal plus background hypothesis. The significance $z$ is then determined as the value at which the integral of a Gaussian between $z$ and $\infty$ results in a value equal to the local p-value. 

\begin{figure}
 \begin{center} 
 \includegraphics[width=0.5\textwidth, height=0.35\textheight]{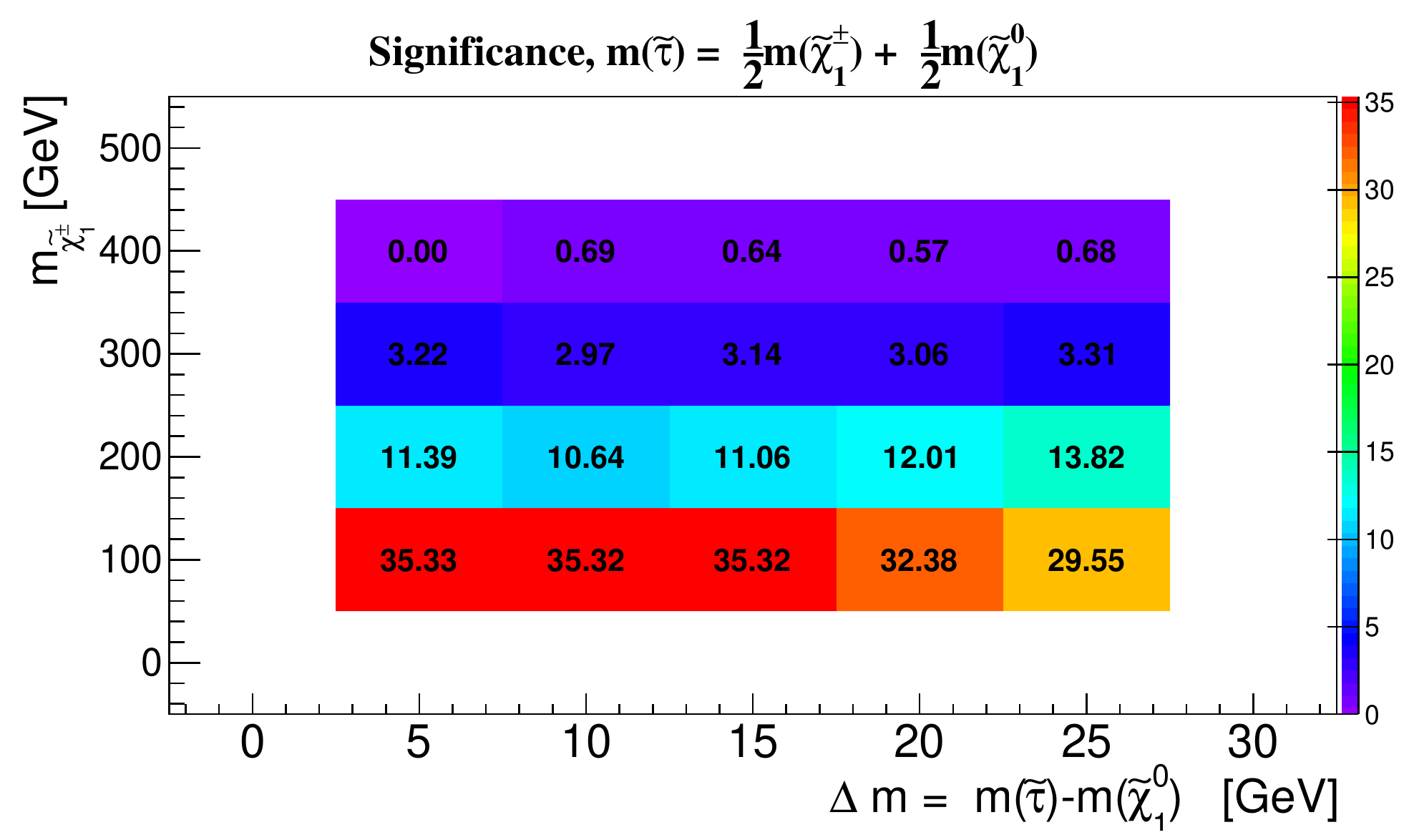}
 \end{center}
 \caption{Signal significance, using a shape based statistical analysis of the $m_{T}$ distribution, as a function of $\tilde{\chi}_{1}^{\pm}$ mass and $m(\tilde{\tau})-m(\tilde{\chi}_{1}^{0})$. No systematic effects have been considered.}
 \label{fig:results}
 \end{figure}

\begin{figure}
 \begin{center} 
 \includegraphics[width=0.5\textwidth, height=0.35\textheight]{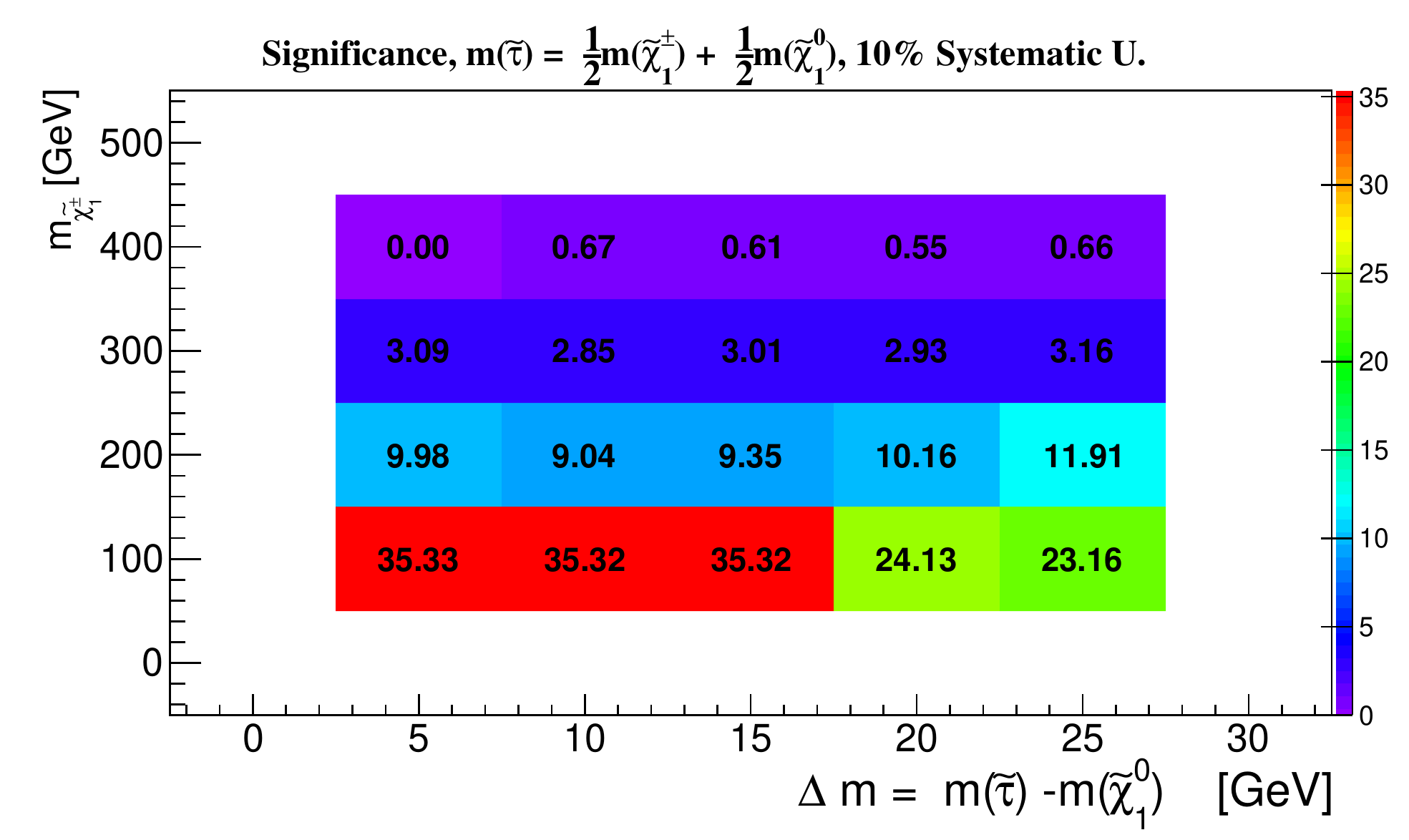}
 \end{center}
 \caption{Signal significance, using a shape based statistical analysis of the $m_{T}$ distribution, as a function of $\tilde{\chi}_{1}^{\pm}$ mass and $m(\tilde{\tau})-m(\tilde{\chi}_{1}^{0})$. A flat systematic effect of 10\%, completely correlated across $m_{T}$ bins, has been considered on the signal and background yields.}
 \label{fig:results1}
 \end{figure}

Figure \ref{fig:results} shows the expected signal significance without considering any systematic effects.  Figure \ref{fig:results1} shows the expected signal significance after considering a flat 10\% systematic effect, completely correlated across $m_{T}$ bins, in the signal and background yields. The proposed methodology can provide $5\sigma$ ($3\sigma$) significance for $\tilde{\chi}_{1}^{\pm}$ masses up to approximately 250 GeV (300 GeV) and with $m(\tilde{\tau})-m(\tilde{\chi}_{1}^{0}) < 25$ GeV, allowing the ATLAS and CMS experiments to probe previously unreachable parts of the $\tilde{\tau}-\tilde{\chi}_{1}^{0}$ co-annihilation phase space important to the connection between particle physics and cosmology. The assumption of a completely correlated systematic uncertainty with respect to $m_{T}$ is based on the belief that the $\tau_{h}$ identification and $p_{T}^{miss}$ trigger efficiencies do not depend on the value of $m_{T}$. This assumption depends on the performance of the improved and updated reconstruction algorithms of the CMS and ATLAS experiments under future running conditions, which is outside the scope of this paper. However, for the luminosity considered, the conclusions have been tested to be independent of the assumption of a completely correlated systematic uncertainty with $m_{T}$. 

\begin{figure}
 \begin{center} 
 \includegraphics[width=0.5\textwidth, height=0.35\textheight]{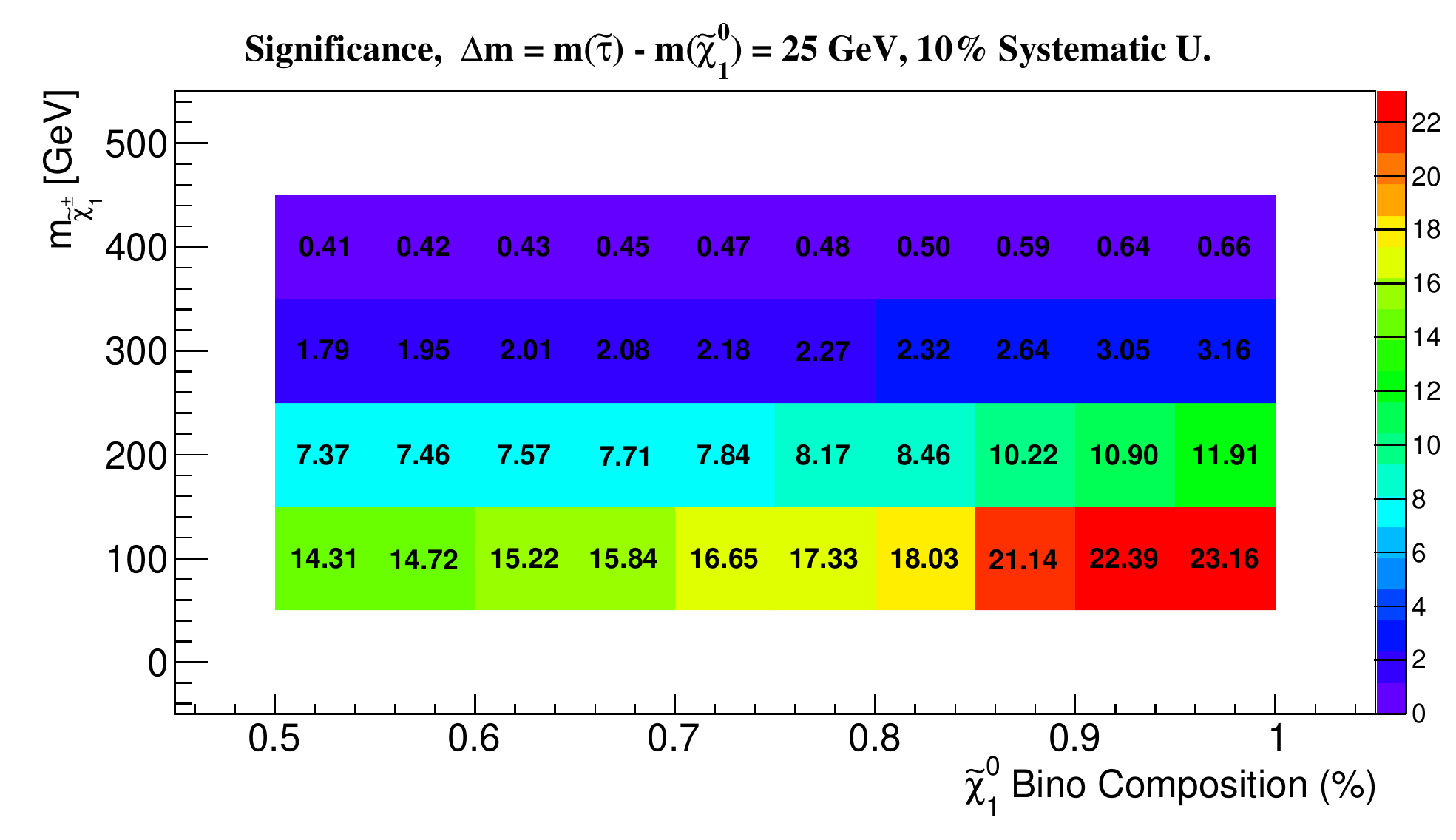}
 \end{center}
 \caption{Signal significance, using a shape based statistical analysis of the $m_{T}$ distribution, as a function of $\tilde{\chi}_{1}^{\pm}$ mass and the fraction of Bino composition of the LSP, for the scenario with $m(\tilde{\tau})-m(\tilde{\chi}_{1}^{0}) < 25$ GeV. A flat systematic effect of 10\%, completely correlated across $m_{T}$ bins, has been considered on the signal and background yields.}
 \label{fig:results3}
 \end{figure}

\begin{figure}
 \begin{center} 
 \includegraphics[width=0.5\textwidth, height=0.35\textheight]{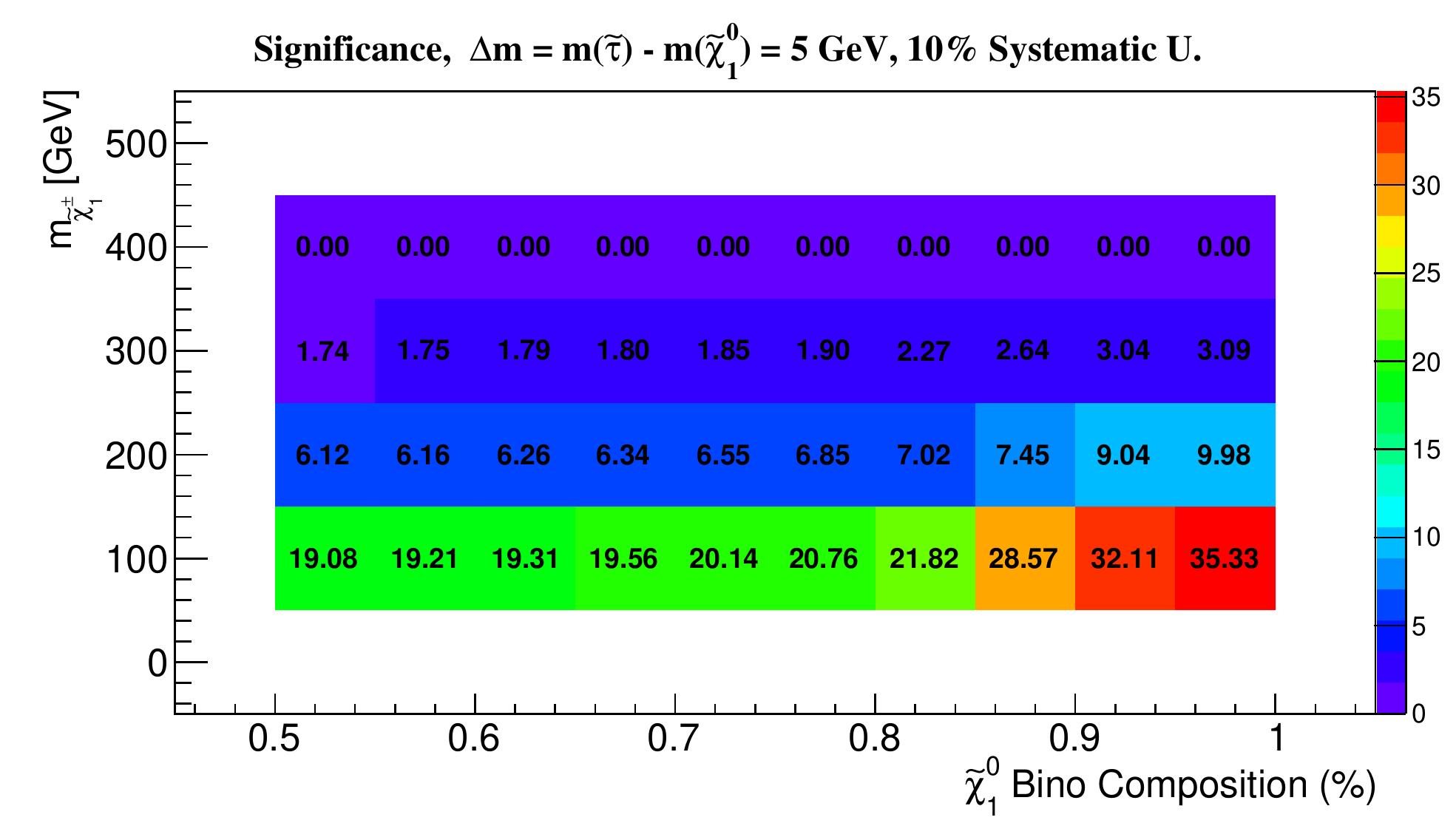}
 \end{center}
 \caption{Signal significance, using a shape based statistical analysis of the $m_{T}$ distribution, as a function of $\tilde{\chi}_{1}^{\pm}$ mass and the fraction of Bino composition of the LSP, for the scenario with $m(\tilde{\tau})-m(\tilde{\chi}_{1}^{0}) < 5$ GeV . A flat systematic effect of 10\%, completely correlated across $m_{T}$ bins, has been considered on the signal and background yields.}
 \label{fig:results2}
 \end{figure}

Although the benchmark signal samples considered thus far focus on the case where the $\tilde{\chi}_{1}^{\pm}$/$\tilde{\chi}_{2}^{0}$ is mostly Wino and the LSP is mostly Bino 
(when co-annihilation can give rise to the correct LSP DM relic density), a study is also performed on the impact of Wino and Bino compositions of the 
$\tilde{\chi}_{1}^{\pm}$/$\tilde{\chi}_{2}^{0}$ and LSP, respectively, to the signal sensitivity. This allows for a more general overview of the impact of the proposed search to compressed SUSY, 
independent of the connection to cosmological DM. For this purpose, signal samples were produced by fixing the $\tilde{\chi}_{1}^{\pm}$/$\tilde{\chi}_{2}^{0}$ and LSP masses and varying 
the $\mu$ parameter, which controls the gaugino mixing. For example, for $m(\tilde{\chi}_{1}^{\pm}) = 100$ GeV and $m(\tilde{\chi}_{1}^{0}) = 50$ GeV, the $\mu$ parameter was 
decreased to produce LSP Bino compositions ranging from 50\% to 97\%. Decreasing the $\mu$ parameter in order to decrease the LSP Bino composition makes the Higgsinos more 
important and thus simultaneously decreases the Wino composition for $\tilde{\chi}_{1}^{\pm}$/$\tilde{\chi}_{2}^{0}$ (i.e. they are no longer mostly wino-like). 
The Wino compositions for $\tilde{\chi}_{1}^{\pm}$/$\tilde{\chi}_{2}^{0}$ range from $\approx 40$\% to 99\%. Figures \ref{fig:results2} and \ref{fig:results3} show the expected signal significance,  
using an integrated luminosity of 30 fb$^{-1}$, as a function of $m(\tilde{\chi}_{1}^{\pm})$ and LSP Bino composition for fixed $\Delta m$ of 25 GeV and 5 GeV respectively. 
For a fixed set of masses, the predicted signal yields decrease as the LSP Bino and $\tilde{\chi}_{1}^{\pm}$/$\tilde{\chi}_{2}^{0}$ Wino compositions decrease, resulting in $\approx$ 55\% 
decrease in signal significance for a LSP Bino composition of 50\%. The signal significances shown in Figures  \ref{fig:results2}  and  \ref{fig:results3}  were calculated using the same statistical procedure 
outlined above and similarly considering a 10\% systematic uncertainty.

\section{Discussion}

The main result of this paper is that the $\tilde{\tau}$-$\tilde{\chi}_{1}^{0}$ co-annihilation region with $\Delta m < 50$ GeV, where experimental sensitivity is limited from current searches performed at the LHC, can be probed using a search strategy of one soft hadronically decaying tau lepton and large missing transverse energy recoiling against a hard $p_{T}$ jet from initial state radiation. These regions of SUSY also play a decisive role in thermal Bino DM cosmology models which require $\tilde{\tau}$-$\tilde{\chi}_{1}^{0}$ co-annihilation to obtain the correct relic DM density observed today. A major highlight of the proposed search strategy is the ability to select low $p_{T}$ hadronic tau decays, facilitated by the use of $p_{T}^{miss}$ triggers from the boost effect of the high $p_{T}$ ISR jet, in order to maximize signal acceptance in these compressed scenarios while simultaneously providing large reduction against SM backgrounds. The ability of the ATLAS and CMS experiments to provide good $\tau_{h}$ identification at low $p_{T}$ is a key ingredient. We find that for $m(\tilde{\tau}) - m(\tilde{\chi}_{1}^{0}) < 25$ GeV, gaugino masses up to 300 GeV (250 GeV) can be probed at 3$\sigma$ (5$\sigma$) level with 30 fb$^{-1}$ of 13 TeV data from the LHC. We emphasize that the experimental constraints for the SUSY parameter space with $m(\tilde{\tau}) - m(\tilde{\chi}_{1}^{0}) < 25$  GeV with the ATLAS and CMS data to date do not exceed those of the LEP experiments, and thus the proposed new search can nicely complement the current analyses performed at the LHC. 

\section{Acknowledgements}

We thank the constant and enduring financial support received for this project from the faculty of science at Universidad de los Andes (Bogot\'a, Colombia), the administrative department of science, technology and innovation of Colombia (COLCIENCIAS), the Physics \& Astronomy department at Vanderbilt University and the US National Science Foundation. This work is supported in part by NSF Award PHY-1506406. We would like to thank Kuver Sinha for useful discussions.

\end{document}